# Optical characterization of deuterated silicon-rich nitride waveguides

Xavier X. Chia,[1] George F. R. Chen,[1] Yanmei Cao,[1] Peng Xing,[1] Doris K. T. Ng,[2] and Dawn T. H. Tan[1,2*]

[1]*Photonics Devices and Systems Group, Engineering Product Development, Singapore University of Technology and Design, 8 Somapah Road, Singapore 487372*
[2]*Institute of Microelectronics, Agency for Science, Technology and Research (A*STAR), 2 Fusionopolis Way, Singapore 138634*
dawn_tan@sutd.edu.sg

**Abstract:** Chemical vapor deposition-based growth techniques allow flexible design of CMOS-compatible materials. Here, we report the deuterated silicon-rich nitride films grown using plasma-enhanced chemical vapor deposition. The linear and nonlinear properties of the films are characterized. We compare the absorption at 1550nm wavelength region for films grown with $SiH_4$ and $SiD_4$, and experimentally confirm that the silicon-rich nitride films grown with $SiD_4$ eliminates Si-H related absorption. Waveguides fabricated on the films are further shown to possess a linear and nonlinear refractive index of 2.46 and $9.8 \times 10^{-18}$ $m^2W^{-1}$ respectively.

## 1. Introduction

Chemical vapor deposition (CVD) tools allow flexible growth of thin films. Through a judicious selection of precursor gases, flow rates and process temperatures, films with varied optical properties may be designed and realized. Amongst the spectrum of materials which may be grown using CVD techniques, silicon and silicon-nitride based materials are of particular interest for photonics and electronics. These materials are compatible with complementary metal oxide semiconductor processes and used in a plethora of applications. Amorphous silicon is a commonly used material for solar cells [1] whereas silicon nitride is a commonly used passivation layer in electronics [2]. With regards to photonics, CVD grown silicon and silicon-nitride based materials possess favorable optical properties such as high transmissivity at the visible (silicon nitride), short-wave infrared (silicon and silicon nitride) to the mid-infrared (silicon and silicon nitride) wavelengths [3-5], importantly also being compatible with complementary metal-oxide semiconductor (CMOS) processing. Silicon-rich nitride is a material whose properties may be engineered to possess the optical properties of both Si and $Si_3N_4$. Indeed, significant progress has been made in silicon-rich nitride based nonlinear optics in recent years, largely due to the high nonlinear refractive index and absence of two-photon absorption at the telecommunications wavelengths.

Silicon-rich nitride films require precursor gases containing silicon and nitride atoms for their growth. Commonly used precursor gases supplying the silicon content are silane ($SiH_4$) and dicholorosilane ($SiH_2Cl_2$), whereas nitrogen ($N_2$) gas and ammonia ($NH_3$) supply the nitrogen content. From the standpoint of photonics applications, $N_2$ is a superior precursor gas than $NH_3$, as it precludes formation of N-H bonds which also possess absorption overtones at the short-wave infrared wavelength [6]. In our prior work, we have aimed to eliminate N-H bonds in ultra-silicon-rich-nitride films by utilizing $N_2$ gas instead of $NH_3$ gas [7-9]. The formation of Si-H bonds in films however is not easily resolved whether $SiH_4$ or $SiH_2Cl_2$ precursor gases are used for the film growth. Consequently, Si-H absorption overtones at the shortwave infrared are a source of absorption in photonic devices [10]. To overcome Si-H absorption, the $Si_3N_4$ community routinely anneals their films at temperatures up to 1200°C [11, 12]. This allows ultra-low loss films to be created. An alternative method of reducing Si-

H related loss involves using precursor gases which do not contain any Si-H bonds to begin with. Deuterated silane (SiD$_4$) is chemically almost identical to SiH$_4$, with the exception that the H atoms are replaced by deuterium, an isotope of hydrogen. Si-D bonds have been reported to have an absorption overtone close to 2μm as opposed to close to 1.55μm in the case of Si-H [13]. Consequently, growth of films using SiD$_4$ allows an elegant and straightforward solution to eliminating Si-H related absorption. Deuterated silane has been used with significant success in Hydex glass, which has spawned a multitude of nonlinear optics advancements over the years [14-16], as well as in the growth of SiO$_2$, SiON and Si$_3$N$_4$ [13, 17]. Deuterium-passivation using D$_2$ gas has also previously been shown to improve the stability of amorphous silicon films [18]. In a similar vein, O-H bonds which form during optical fiber manufacturing have historically posed a problem due to their absorption being located at the O-band. The replacement of O-H bonds with O-D bonds has also been shown to allow significant loss reductions at the O-band [19]. In practice, the optical fiber is exposed to an atmosphere of deuterium during the manufacturing process to reduce the hydrogen related attenuation [20]. The use of deuterium is therefore advantageous in many respects for photonics applications.

In this manuscript, we report the growth and optical characterization of silicon-rich nitride films grown using SiD$_4$ gas (SRN:D). Materials characterization using Fourier Transform Infrared Spectroscopy verify the absence of Si-H bonds in the films grown using plasma enhanced chemical vapor deposition. Waveguide devices are fabricated using the films and the linear and nonlinear properties are reported. We achieve a nonlinear refractive index of 9.8×10$^{-18}$ m$^2$W$^{-1}$, an order of magnitude larger than in stoichiometric silicon nitride, and a waveguide nonlinear parameter of 98 W$^{-1}$m$^{-1}$. The films are further characterized to have a bandgap of 1.94 eV, indicating that two-photon absorption is absent at a wavelength of 1550 nm.

## 2. Deuterated Silicon-Rich Nitride Materials Growth and Characterization

The fundamental vibrational absorption for Si-H bonds has been reported to occur at the wavenumber region at around 2200 cm$^{-1}$ [21-25]. The third vibrational overtone which resides at the region around 6600 cm$^{-1}$ (1.5 μm wavelength region) is responsible for the deleterious attenuation associated with Si-H bonds at telecommunications wavelengths. The quantum harmonic oscillator model may be used to describe the vibrational bond energies associated with Si-H bonds [26]:

$$-\frac{h^2}{2m}\frac{d^2\psi}{dx^2} + \frac{1}{2}kx^2\psi = E\psi \qquad (1)$$

where $h$ is Planck's constant, $\Psi$ is the wave function, $x$ is the displacement from equilibrium and $E$ is energy. Enforcing the boundary conditions, $\Psi = 0$, $x = \pm \infty$, generates the closed form expression for the vibrational bond energy as given by [26]:

$$E_v = h.(1/2\pi).(v + 1/2)\sqrt{k/m_1 + k/m_2} \qquad (2)$$

where $v$ is the vibrational quantum number (an integer), $k$ is the bond's force constant and $m_{1,2}$ is the atomic mass for the two atoms making up the bond. Referring to the fundamental vibrational absorption for Si-H bonds in the context of Eq. (2), it may be observed that a change in m$_1$ or m$_2$ will result in a change in the vibrational bond energy. Given that the atomic mass of deuterium (D), hydrogen (H) and silicon (Si) is 1, 2 and 28 respectively, the fundamental vibrational bond energy for Si-H is expected to decrease with the substitution of H with the D atom. From Eq. (2), it may be derived that Si-H's third vibrational absorption overtone which resides at the region around 6600 cm$^{-1}$ (1.5 μm region) will be reduced to 4670 cm$^{-1}$ (2.1 μm region). Consequently, the replacement of Si-H bonds with Si-D bonds by using deuterated silane instead of silane gas generates a significant red-shift in the Si-H absorption associated with the vibrational bond energy at the telecommunications wavelength region, such that any material related absorption arising from Si-H will be eliminated.

The SRN: D films are grown using plasma enhanced chemical vapor deposition at a temperature of 350°C. To fabricate the SRN:D films, deuterated silane (SiD$_4$) and N$_2$ gas were used in lieu of conventional silane (SiH$_4$) and NH$_3$. In the deposition, we utilized a SiD$_4$ and N$_2$ flow rates with a ratio of 3:250. Films with a thickness of 310 nm were then deposited on a silicon substrate with a 5 μm thermal SiO$_2$ layer. The use of deuterated silane and N$_2$ replaces the Si-H and N-H bonds with Si-D and N-D respectively, resulting in the shifting of bond absorbance overtones away from the telecommunications region at 1.55 μm [6, 13, 25].

We first characterize the material properties of the SRN:D films. Fourier-Transform Infrared (FTIR) spectroscopy was conducted to determine the material absorption of the deposited films (Fig. 1a). Regions where Si-H bonds and N-H bonds could occur [21-24] are highlighted at wavenumber range 2157 cm$^{-1}$ to 2250 cm$^{-1}$ and 3290 cm$^{-1}$ to 3464 cm$^{-1}$ respectively. We note that there are no peak occurrences in these regions, indicating that Si-H and N-H absorption bonds are absent or negligible. In conventional silicon-rich nitride films using SiH$_4$-based gas precursors, peaks are easily observed in the absorbance spectra corresponding to Si-H and N-H bonds [21, 23], evidencing the presence of Si-H and N-H absorption bonds. We further investigate the FTIR data, specific at Si-H bond and N-H bond regions to quantify the Si-H and N-H bonds. Small peaks are identified at wavenumbers ~2200 cm$^{-1}$ and ~3350 cm$^{-1}$ where Si-H and N-H bonds are likely to occur [24]. We quantify the bond concentrations by relating the areas of the Si-H and N-H absorbance lines with the concentrations of Si-H and N-H bonds using Lanford and Rand's technique [27], with equations [21, 24, 27]:

$$[Si - H] = \frac{1}{2.303 \times \sigma_{Si-H}} \times \int_{band} \alpha(\omega)d\omega \quad (3)$$

$$[N - H] = \frac{1}{2.303 \times \sigma_{N-H}} \times \int_{band} \alpha(\omega)d\omega \quad (4)$$

Here, $\sigma_{Si-H} = 7.4 \times 10^{-18} cm^2$ is the absorption cross-section of Si-H and $\sigma_{N-H} = 5.3 \times 10^{-18} cm^2$ is the absorption cross-section of N-H, $\int \alpha(\omega)d\omega$ is the baseline absorption area of the band. In addition,

$$\alpha = \frac{2.303}{t} A \quad (5)$$

Here, $A$ is the absorbance and $t$ is the film thickness. Si-H bond concentration is calculated to be $1.17 \times 10^{20} \, cm^{-3}$ and N-H bond concentration is calculated to be $9.36 \times 10^{19} \, cm^{-3}$. Compared to Si-H and N-H bond concentrations reported in literature [21, 24] using SiH$_4$-based gas precursors which calculate the bond concentrations to be in the range of $10^{21}$-$10^{22} \, cm^{-3}$, the Si-H and N-H bond concentrations in our SRN:D films are 1-2 orders of magnitude smaller, further confirming the effectiveness of the SiD$_4$ gas in significantly reducing Si-H bonds.

The $n$ and $k$ data were also extracted (Fig. 1b), showing a linear refractive index of 2.46 at a wavelength of 1550 nm, following which the material dispersion for the film was calculated (Fig. 1c) [28] and it is observed that the material dispersion is normal at the 1550nm wavelength region.

Subsequently, Tauc's Method [29] was applied to determine an optical bandgap of 1.94 eV. Tauc's Method is derived from the governing expression relating absorption coefficient of an amorphous material, $\alpha$ with its frequency and bandgap [30]:

$$(hf\alpha)^{\frac{1}{2}} = B(hf - E_g) \quad (6)$$

Here, $h$ is Planck's constant, $f$ is the frequency of the incident photon, $r$ is the nature of the electron transition, $B$ is a material-dependent constant, and $E_g$ is the bandgap of the material.

The optical bandgap of the material can be extracted by plotting $(hf\alpha)^{\frac{1}{2}}$ vs $hf$ and extrapolating a fit of the linear portion of the plot to determine the x-intercept. Fig 1d shows the Tauc plot from the 310 nm SRN:D film and the optical bandgap is determined to be 1.94 eV.

The optical bandgap of a medium determines the wavelengths at which non-negligible material losses are present in the film [31]. A bandgap of 1.94 eV corresponds to a wavelength of 640 nm, indicating that the SRN-D film is well suited for applications both within the telecommunications bands at 1550 nm. An optical bandgap of 1.94 eV also eliminates the effects of two-photon absorption (TPA) around 1550 nm which can be observed amorphous silicon devices, which makes SRN:D a suitable media for nonlinear applications in the telecommunications band.

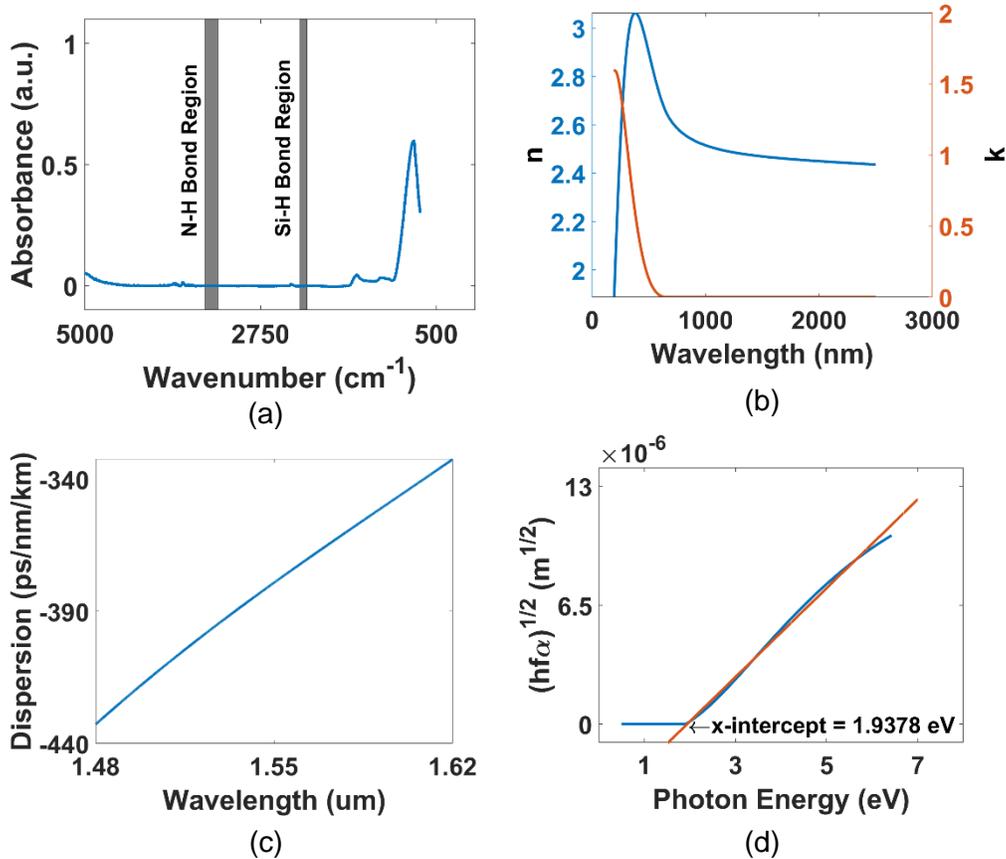

Fig. 1. (a) Baseline-adjusted absorbance plot exhibiting regions where Si-H and N-H bonds could occur. We note no peak occurrences in these regions, indicating absence or negligible Si-H and N-H bonds. (b) Refractive index n (—) and extinction coefficient k (—) data extracted using FTIR measurements (c) Material dispersion curves from 1480 nm-1620 nm for the SRN-D film (d) Tauc's plot of the SRN-D film with the original curve (—) and fitted line (—)

The characterisation of linear optical properties of the film necessitates the fabrication of waveguides and resonators. A series of waveguides, microring, and racetrack resonators were patterned using electron-beam lithography and etched with reactive ion etching, following which a 2 μm thick layer of SiO$_2$ was deposited to form the upper cladding. Figure 2a shows a

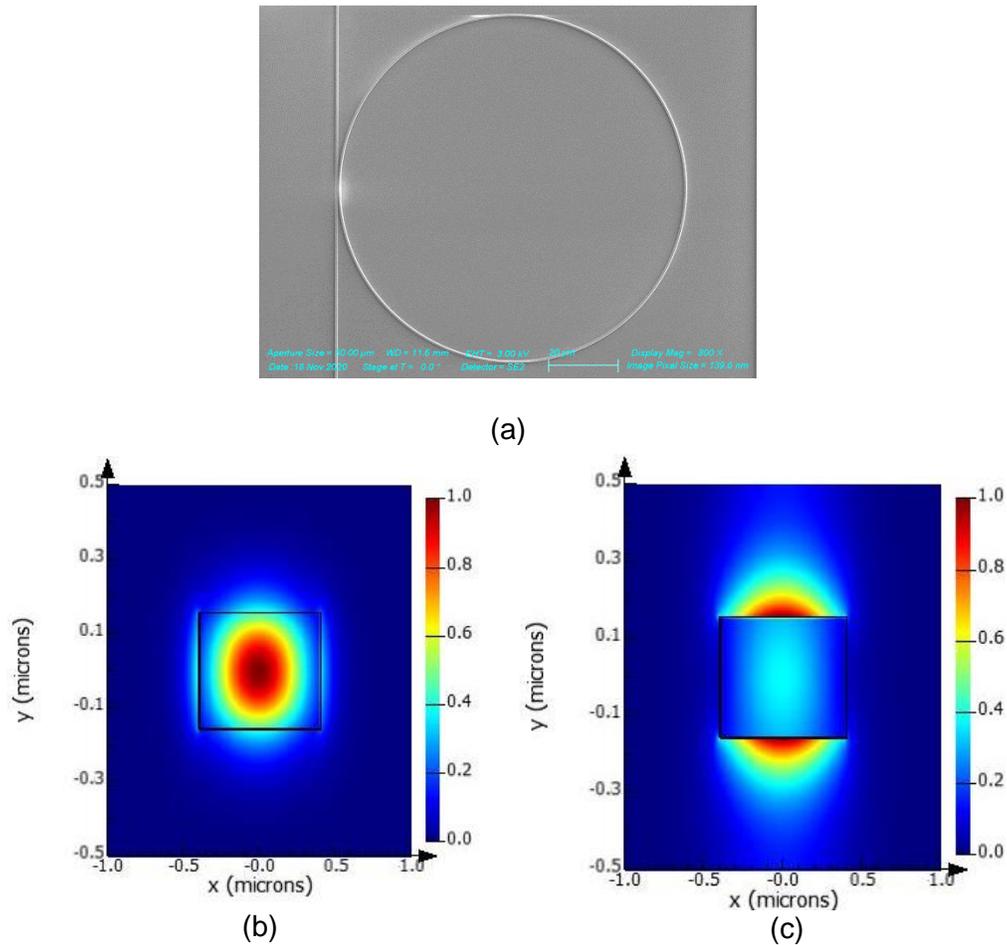

Fig. 2 (a) SEM Micrograph of all-pass ring resonator (b) Mode Profile of Quasi-TE mode at 1.55 µm (c) Mode Profile of quasi-TM Mode at 1.55 µm scanning electron micrograph of a fabricated ring resonator. Figure 2b and c further show the mode profiles for the quasi-TE and quasi-TM modes of the waveguide.

A broadband tunable single-mode continuous wave laser source was edge-coupled using polarization-maintaining fibers into the devices. To determine the propagation loss for each waveguide width, cutback measurements were conducted and the results are shown in Figure 3a. In both TE and TM modes, the propagation loss is observed to decrease with an increase in waveguide width across the same film thickness, indicating that sidewall roughness is a key contributor to propagation loss given the increased mode confinement in the smaller waveguide widths studied [32]. The source was similarly coupled into the resonators, following which a benchmark of attainable intrinsic quality factor and the group index of these waveguides could be determined. Figure 3b shows the transmission spectrum of the microring resonator of radius 50 µm and gap 300 nm, when quasi-TE light is coupled into the device.

## 3. Experimental Characterization

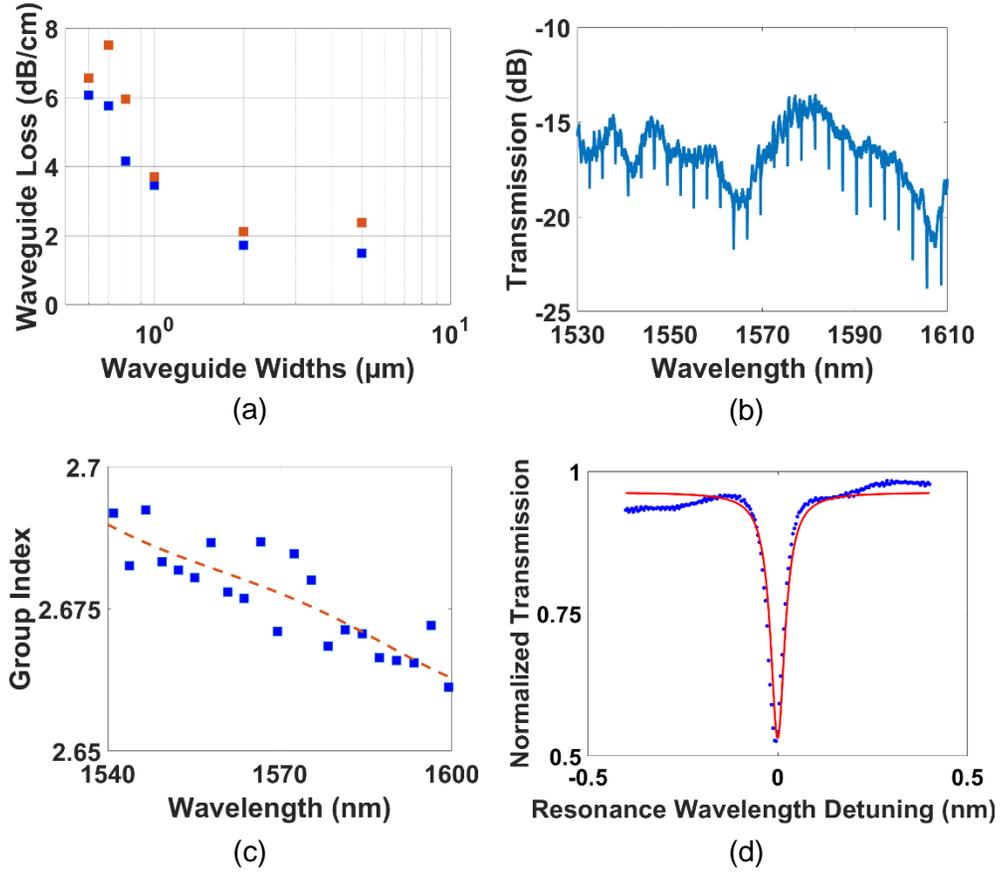

Fig. 3. (a) Waveguide propagation losses from cutback measurements for TE (■) and TM (■) Modes (b) Transmission spectrum for microring resonator of radius 50 µm and gap 300 nm for the quasi-TE mode. (c) Experimental (■) and fitted (--) Group Index extracted from microring resonator of radius 50 µm and gap 300 nm (d) Sample Lorentzian fit showing experimental data of a resonant dip (∘) and fitted curve (—).

The free spectral range (FSR) of the resonators can be obtained as follows [33]:

$$\text{FSR} = \frac{\lambda_0^2}{n_g L} \quad (7)$$

Here, $\lambda_0$ is the resonant wavelength, $n_g$ the group index, and $L$ the optical path-length of the resonator. From the transmission spectrum of the devices (Figure 3b), the FSR at the resonances may be obtained by identifying the wavelength differences between each resonant dip and the group index of the devices experimentally obtained (Figure 3c).

The loaded ($Q_L$) and intrinsic ($Q_{int}$) quality factors can also be obtained [34]:

$$Q_L = \frac{\lambda_0}{\text{FWHM}} \quad (8)$$

$$Q_{int} = \frac{2Q_L}{1+\sqrt{T_0}} \tag{9}$$

Here, FWHM is the full-width half-maximum of the resonant dips when fitted to a Lorentzian (Fig. 3d) and $T_0$ is the fractional transmission at the resonance. Of the fabricated devices, a microring resonator of waveguide width 0.85 µm and gap 0.3 µm yielded the highest loaded quality factor of 62,000, corresponding to an intrinsic quality factor of 64,700. In general, the quality factors of the ring resonators decrease with increasing wavelength, suggesting that the effective index of the devices are also lower at higher wavelengths.

The waveguide dispersion of the devices can be determined by extracting the group index of the resonators from the FSR [28]:

$$\beta_2 = \frac{1}{c} \cdot \frac{dn_g}{d\omega} \tag{10}$$

$$D = -\frac{2\pi c}{\lambda^2}\beta_2 = -\frac{2\pi}{\lambda^2}\left(\frac{dn_g}{d\omega}\right) = \frac{1}{c} \cdot \frac{dn_g}{d\lambda} \tag{11}$$

Here, $c$ is the speed of light in a vacuum, $\omega$ is the angular frequency, $D$ is the dispersion in ps/nm/km, and $\beta_2$ is the 2nd order group velocity dispersion. The decreasing group index as a function of wavelength suggests that the dispersion of the fabricated waveguide shown in Fig. 3 (c) is normal, which is expected for lower film thicknesses. Anomalous dispersion is an important property for nonlinear optics phenomena such as efficient four-wave mixing and frequency comb generation, and while the fabricated films exhibit normal dispersion it is possible to engineer the waveguide geometry to bring about anomalous dispersion.

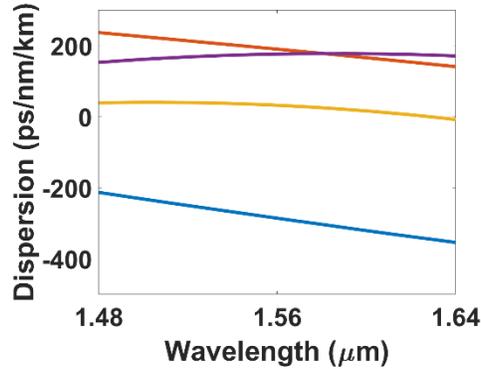

Fig. 4 Simulated waveguide dispersion for waveguide width of 0.8 µm and film thickness 310 nm (—), 400 nm (—), 500 nm (—), and 600 nm (—)

Numerical simulations of various waveguide geometries were conducted to study the regime in which anomalous dispersion can be achieved. Figure 4 shows the results of these calculations and results suggest that the thickness of the SRN:D film needs to be increased beyond 500nm to approach the required conditions for anomalous dispersion for the wavelength region of 1.48 µm to 1.64 µm.

The experimentally determined linear parameters such as the propagation loss, group indices, and dispersion of the devices also allow us to determine nonlinear parameters of the platform and the fabricated devices, providing insight into the suitability of the material as a potential platform for nonlinear optics. To characterize the nonlinearity of the waveguides, self-phase modulation (SPM) experiments were conducted. The maximum nonlinear phase-shift of pulses undergoing SPM ($\phi_{NL}$) is dependent on the nonlinear parameter ($\gamma$) of the waveguides and the effective length ($L_{eff}$). The expression for these governing parameters are given by [28]:

$$\phi_{NL} = \gamma P_0 L_{eff} \tag{12}$$

$$L_{eff} = \frac{[1 - \exp(-\alpha L)]}{\alpha} \tag{13}$$

$$\gamma = \frac{2\pi n_2}{\lambda A_{eff}} \tag{14}$$

Here, $P_0$ is the peak input power, $L$ is the waveguide length in cm, $\alpha$ is the propagation loss in dB/cm, $A_{eff}$ is the effective modal area, and $n_2$ is the Kerr nonlinearity of the film. The SPM experiments were conducted by coupling 1.3 ps pulses from a fiber laser at a repetition rate of 20 MHz. Waveguides of width 0.8 µm and length 6.9 mm were chosen for their high modal confinement and lower propagation loss, and the input power was varied by displacing the input fiber. Figure 3a shows the evolution of the pulse spectrum as the input power is increased, reaching a maximum observed phase shift of 0.8π at a peak pulse power of 5.9 W.

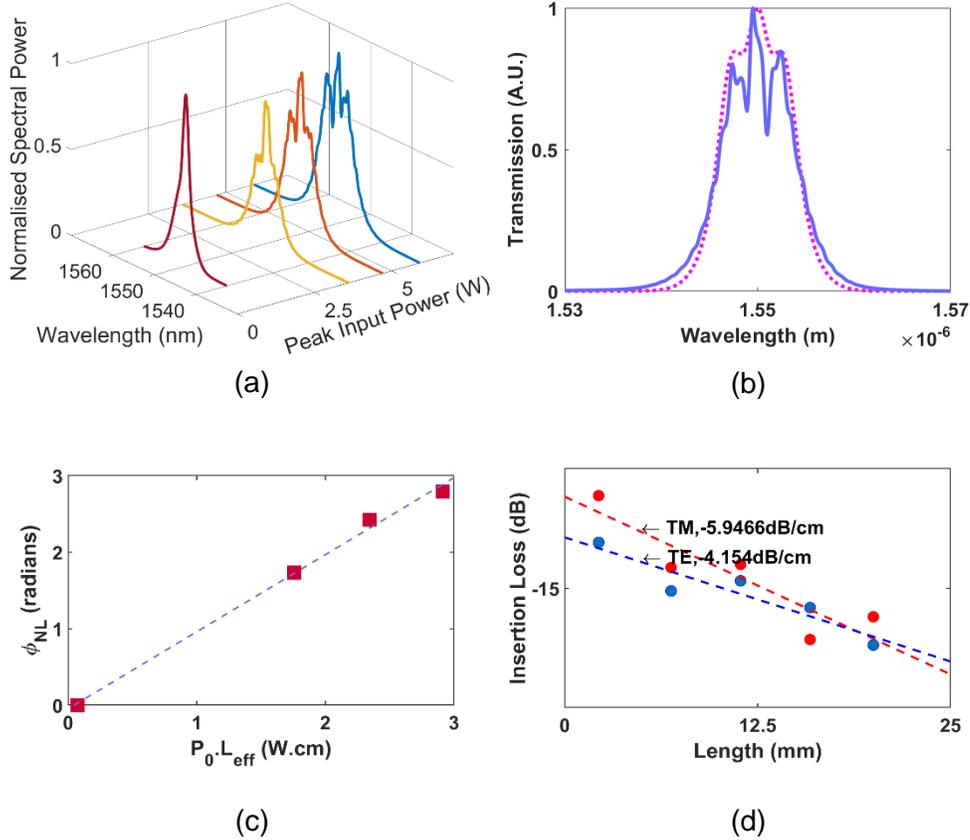

Fig. 5 (a) SPM spectrum plots for peak input powers 3.6 W (—), 4.7 W (—), 5.9 W (—), and Source (—) (b) Modelled (···) and experimentally measured (—) SPM spectra for $P_0$ = 5.9 W. (c) Parameter fit for experimental data (■) and fitted line (--). (d) Cutback measurements for the waveguide used for the SPM measurements, for both the TE (●) and TM mode (●).

The nonlinear parameter $\gamma$ is extracted by plotting the SPM phase shift against the peak input power. Figure 5c plots the nonlinear phase shift as a function of $P_0 L_{eff}$ where a linear relationship is obtained. From Eq. (12), the gradient of the plot in Fig. 5c allows us to extract a nonlinear parameter of $\gamma = 95\ \text{W}^{-1}\text{m}^{-1}$ for the waveguide. For a waveguide of 0.8 µm × 0.31 µm at a wavelength of 1.55 µm, the effective modal area $A_{eff}$ was determined to be 0.42 µm², following which the Kerr nonlinearity was determined using Eq. (14) to be $9.8 \times 10^{-18}\ \text{m}^2\text{W}^{-1}$. We model the SPM dynamics of the pulse propagation in the waveguide in order to corroborate the experimentally extracted nonlinear parameter. The pulse propagation dynamics may be described using the nonlinear Schrödinger equation [24]:

$$\frac{\partial A}{\partial z} = -\frac{\alpha}{2}A - i\frac{\beta_2}{2}\frac{\partial^2 A}{\partial t^2} + i\gamma|A|^2 A \quad (15)$$

where $z$ is the longitudinal coordinate, $t$ is time, $\alpha$ is the linear loss coefficient, $A$ and $\omega_0$ are the slow varying pulse envelope and carrier frequency, respectively. Figure 5 (b) shows the modelled and experimental pulse spectrum at an input peak power of 5.9 W, where good agreement is achieved. Referring to Fig. 5 (d) where the insertion loss for the waveguides is plotted for both the quasi-TE and quasi-TM modes and the propagation losses are extracted to be 4.2 dB/cm and 5.9dB/cm respectively. The relationship between waveguide width and propagation losses (Fig. 2 (a)) show that wider waveguides have lower losses, implying that the surface roughness from the fabrication is the dominant contributor to the propagation losses.

## 4. Conclusion

We have fabricated silicon-rich nitride films using $SiD_4$ and $N_2$ precursor gases, allowing films to be grown without Si-H and N-H bonds, both of which are typically present in films fabricated using $SiH_4$, $SiH_2Cl_2$ and/or $NH_3$ gas. Consequently, bond overtones normally present at the telecommunications bands are absent, minimizing a key source of optical absorption and rendering the platform ideal for integrated optics applications around 1550 nm. Using FTIR measurements, we confirm the absence of Si-H and N-H bonds within the films.

The experimentally determined $\gamma$ and $n_2$ of the fabricated waveguides, $95\ \text{W}^{-1}\text{m}^{-1}$ and $9.8 \times 10^{-18}\ \text{m}^2\text{W}^{-1}$ respectively, a value which is close to that reported in crystalline silicon [35]. Calculations suggest that anomalous dispersion could be achieved with larger film thicknesses. The optical bandgap of 1.94eV extracted using Tauc's method denotes the absence of two-photon absorption near 1550nm, making SRN:D a promising platform for power efficient nonlinear optical applications. Further optimization of the film could yield lower propagation losses. This coupled with the low inherent nonlinear loss of the films could facilitate the realisation of frequency comb and supercontinuum generation at low peak pulse powers [13, 36]. Further process optimization for achieving lower propagation losses may also be undertaken to optimize the smoothness of the etched surfaces.

In previous work, ultra-silicon-rich-nitride films demonstrated a nonlinear refractive index 100 times greater than stoichiometric silicon nitride [8, 37-40]. Prevailing characterization of silicon-nitride thin films fabricated with different ratios of precursor gases suggests that the linear and nonlinear refractive indices of the SRN:D films increase correspondingly with silicon content [38, 41-44]. Deuterated silicon-rich nitride films with optimized silicon content may therefore be a promising platform for a low-loss, CMOS-compatible, highly nonlinear integrated optics devices, importantly minimizing Si-H and N-H related absorption at the telecommunications bands.

## Acknowledgments

Funding from the National Research Foundation Competitive Research Grant (NRF-CRP18-2017-03), Ministry of Education ACRF Tier 2 Grant and A*STAR grant is gratefully acknowledged.

**References**

[1] J. Kim et al., "10.5% efficient polymer and amorphous silicon hybrid tandem photovoltaic cell," *Nature Communications,* vol. 6, no. 1, p. 6391, 2015/03/04 2015, doi: 10.1038/ncomms7391.

[2] H. Miyazaki, H. Kojima, and K. Hinode, "Passivation effect of silicon nitride against copper diffusion," *Journal of Applied Physics,* vol. 81, no. 12, pp. 7746-7750, 1997/06/15 1997, doi: 10.1063/1.365380.

[3] W. D. Sacher et al., "Visible-light silicon nitride waveguide devices and implantable neurophotonic probes on thinned 200 mm silicon wafers," *Opt. Express,* vol. 27, no. 26, pp. 37400-37418, 2019/12/23 2019, doi: 10.1364/OE.27.037400.

[4] D. J. Blumenthal, R. Heideman, D. Geuzebroek, A. Leinse, and C. Roeloffzen, "Silicon Nitride in Silicon Photonics," *Proceedings of the IEEE,* vol. 106, no. 12, pp. 2209-2231, 2018, doi: 10.1109/JPROC.2018.2861576.

[5] R. Baets et al., "Silicon Photonics: silicon nitride versus silicon-on-insulator," in *Optical Fiber Communication Conference*, Anaheim, California, 2016/03/20 2016: Optical Society of America, in OSA Technical Digest (online), p. Th3J.1, doi: 10.1364/OFC.2016.Th3J.1. [Online]. Available: http://www.osapublishing.org/abstract.cfm?URI=OFC-2016-Th3J.1

[6] A. V. Osinsky, R. A. Bellman, I. A. Akwani, P. A. Sachenik, S. L. Logunov, and J. W. McCamy, "Optical loss mechanisms in GeSiON planar waveguides," *Applied Physics Letters,* vol. 81, no. 11, pp. 2002-2004, 2002/09/09 2002, doi: 10.1063/1.1507611.

[7] D. T. H. Tan et al., "Ultra-Silicon-Rich Nitride Based Devices for High Nonlinear Figure of Merit Photonics Applications," in *2018 Conference on Lasers and Electro-Optics Pacific Rim (CLEO-PR)*, 29 July-3 Aug. 2018 2018, pp. 1-2.

[8] D. T. H. Tan et al., "Nonlinear optics in ultra-silicon-rich nitride devices: recent developments and future outlook," *Advances in Physics: X,* vol. 6, no. 1, p. 1905544, 2021/01/01 2021, doi: 10.1080/23746149.2021.1905544.

[9] B.-U. Sohn, J. W. Choi, D. K. T. Ng, and D. T. H. Tan, "Optical nonlinearities in ultra-silicon-rich nitride characterized using z-scan measurements," *Scientific Reports,* vol. 9, no. 1, p. 10364, 2019/07/17 2019, doi: 10.1038/s41598-019-46865-7.

[10] L. A. Chuprov, P. G. Sennikov, K. G. Tokhadze, S. K. Ignatov, and O. Schrems, "High-resolution Fourier-transform IR spectroscopic determination of impurities in silicon tetrafluoride and silane prepared from it," *Inorganic Materials,* vol. 42, no. 8, pp. 924-931, 2006/08/01 2006, doi: 10.1134/S0020168506080231.

[11] J. C. Bruyère, B. Reynes, C. Savall, and C. Roch, "Annealing of silicon nitride thin films prepared by plasma-enhanced chemical vapor deposition with helium dilution," *Thin Solid Films,* vol. 221, no. 1, pp. 65-71, 1992/12/10/ 1992, doi: https://doi.org/10.1016/0040-6090(92)90797-F.

[12] J. S. Michael, G. Junpeng, V. Gregory Allen, H. Scott, and T. S. Charles, "Fabrication techniques for low-loss silicon nitride waveguides," in *Proc.SPIE*, 2005, vol. 5720. [Online]. Available: https://doi.org/10.1117/12.588828. [Online]. Available: https://doi.org/10.1117/12.588828

[13] J. Chiles et al., "Deuterated silicon nitride photonic devices for broadband optical frequency comb generation," *Opt. Lett.,* vol. 43, no. 7, pp. 1527-1530, 2018/04/01 2018, doi: 10.1364/OL.43.001527.

[14] M. Ferrera et al., "Low-power continuous-wave nonlinear optics in doped silica glass integrated waveguide structures," *Nature Photonics,* vol. 2, no. 12, pp. 737-740, 2008/12/01 2008, doi: 10.1038/nphoton.2008.228.

[15] L. Razzari et al., "CMOS-compatible integrated optical hyper-parametric oscillator," *Nature Photonics,* vol. 4, no. 1, pp. 41-45, 2010/01/01 2010, doi: 10.1038/nphoton.2009.236.

[16] D. J. Moss, R. Morandotti, A. L. Gaeta, and M. Lipson, "New CMOS-compatible platforms based on silicon nitride and Hydex for nonlinear optics," *Nature Photonics,* vol. 7, no. 8, pp. 597-607, 2013/08/01 2013, doi: 10.1038/nphoton.2013.183.

[17] W. Jin, D. D. John, J. F. Bauters, T. Bosch, B. J. Thibeault, and J. E. Bowers, "Deuterated silicon dioxide for heterogeneous integration of ultra-low-loss waveguides," (in eng), *Opt Lett,* vol. 45, no. 12, pp. 3340-3343, Jun 15 2020, doi: 10.1364/ol.394121.

[18] P. Girouard, L. H. Frandsen, M. Galili, and L. K. Oxenløwe, "A Deuterium-Passivated Amorphous Silicon Platform for Stable Integrated Nonlinear Optics," in *Conference on Lasers and Electro-Optics*, San Jose, California, 2018/05/13 2018: Optical Society of America, in OSA Technical Digest (online), p. SW4I.2, doi: 10.1364/CLEO_SI.2018.SW4I.2. [Online]. Available: http://www.osapublishing.org/abstract.cfm?URI=CLEO_SI-2018-SW4I.2

[19] J. Stone, "Reduction of OH Absorption in Optical Fibers by OH → OD Isotope Exchange," *Industrial & Engineering Chemistry Product Research and Development,* vol. 25, no. 4, pp. 609-621, 1986/12/01 1986, doi: 10.1021/i300024a603.

[20] E. Regnier, F. Gooijer, F. Geerings Stephanus Gerardus, E. Burov, A. Bergonzo, and A. Pastouret, "DEUTERIUM TREATMENT METHOD FOR OPTICAL FIBRES," US Patent US 2010/0251775 A1 Patent Appl. US 72860210 A, 2010/10/07, 2010. [Online]. Available: https://lens.org/105-700-061-817-159

[21]　F. Ay and A. Aydinli, "Comparative investigation of hydrogen bonding in silicon based PECVD grown dielectrics for optical waveguides," *Optical Materials,* vol. 26, no. 1, pp. 33-46, 2004/06/01/ 2004, doi: https://doi.org/10.1016/j.optmat.2003.12.004.

[22]　K. T. N. Doris, P. Xing, F. R. C. George, H. Gao, Y. Cao, and T. H. T. Dawn, "Improved CMOS-compatible ultra-silicon-rich nitride for non-linear optics," in *Proc.SPIE*, 2021, vol. 11682, doi: 10.1117/12.2582841. [Online]. Available: https://doi.org/10.1117/12.2582841

[23]　I. Guler, "Optical and structural characterization of silicon nitride thin films deposited by PECVD," *Materials Science and Engineering: B,* vol. 246, pp. 21-26, 2019/07/01/ 2019, doi: https://doi.org/10.1016/j.mseb.2019.05.024.

[24]　S. C. Mao *et al.*, "Low propagation loss SiN optical waveguide prepared by optimal low-hydrogen module," *Opt. Express,* vol. 16, no. 25, pp. 20809-20816, 2008/12/08 2008, doi: 10.1364/OE.16.020809.

[25]　T. Shimanouchi, "Tables of molecular vibrational frequencies. Consolidated volume II," *Journal of Physical and Chemical Reference Data,* vol. 6, no. 3, pp. 993-1102, 1977/07/01 1977, doi: 10.1063/1.555560.

[26]　P. W. Atkins and J. De Paula, *Physical chemistry*. Oxford: Oxford University Press (in English), 2006.

[27]　W. A. Lanford and M. J. Rand, "The hydrogen content of plasma-deposited silicon nitride," *Journal of Applied Physics,* vol. 49, no. 4, pp. 2473-2477, 1978/04/01 1978, doi: 10.1063/1.325095.

[28]　G. Agrawal, "Chater 4 - Self-Phase Modulation," in *Nonlinear Fiber Optics (Fifth Edition)*, G. Agrawal Ed. Boston: Academic Press, 2013, pp. 87-128.

[29]　S. V. Deshpande, E. Gulari, S. W. Brown, and S. C. Rand, "Optical properties of silicon nitride films deposited by hot filament chemical vapor deposition," *Journal of Applied Physics,* vol. 77, no. 12, pp. 6534-6541, 1995, doi: 10.1063/1.359062.

[30]　P. Makuła, M. Pacia, and W. Macyk, "How To Correctly Determine the Band Gap Energy of Modified Semiconductor Photocatalysts Based on UV–Vis Spectra," *The Journal of Physical Chemistry Letters,* vol. 9, no. 23, pp. 6814-6817, 2018/12/06 2018, doi: 10.1021/acs.jpclett.8b02892.

[31]　C. J. Krückel, A. Fülöp, Z. Ye, P. A. Andrekson, and V. Torres-Company, "Optical bandgap engineering in nonlinear silicon nitride waveguides," *Opt. Express,* vol. 25, no. 13, pp. 15370-15380, 2017/06/26 2017, doi: 10.1364/OE.25.015370.

[32]　F. Grillot, L. Vivien, S. Laval, D. Pascal, and E. Cassan, "Size influence on the propagation loss induced by sidewall roughness in ultrasmall SOI waveguides," *IEEE Photonics Technology Letters,* vol. 16, no. 7, pp. 1661-1663, 2004, doi: 10.1109/LPT.2004.828497.

[33]　W. Bogaerts *et al.*, "Silicon microring resonators," *Laser & Photonics Reviews,* vol. 6, no. 1, pp. 47-73, 2012, doi: https://doi.org/10.1002/lpor.201100017.

[34]　P. An *et al.*, "Experimental optimisation of O-ring resonator Q-factor for on-chip spontaneous four wave mixing," *Journal of Physics: Conference Series,* vol. 1124, p. 051047, 2018/12 2018, doi: 10.1088/1742-6596/1124/5/051047.

[35]　Q. Lin, O. J. Painter, and G. P. Agrawal, "Nonlinear optical phenomena in silicon waveguides: Modeling and applications," *Opt. Express,* vol. 15, no. 25, pp. 16604-16644, 2007/12/10 2007, doi: 10.1364/OE.15.016604.

[36]　T. Wang *et al.*, "Supercontinuum generation in bandgap engineered, back-end CMOS compatible silicon rich nitride waveguides," *Laser & Photonics Reviews,* https://doi.org/10.1002/lpor.201500054 vol. 9, no. 5, pp. 498-506, 2015/09/01 2015, doi: https://doi.org/10.1002/lpor.201500054.

[37]　E. Sahin, D. K. T. Ng, and D. T. H. Tan, "Optical parametric gain in CMOS-compatible sub-100 μm photonic crystal waveguides," *APL Photonics,* vol. 5, no. 6, p. 066108, 2020/06/01 2020, doi: 10.1063/5.0003633.

[38]　D. K. T. Ng *et al.*, "Exploring High Refractive Index Silicon-Rich Nitride Films by Low-Temperature Inductively Coupled Plasma Chemical Vapor Deposition and Applications for Integrated Waveguides," *ACS Applied Materials & Interfaces,* vol. 7, no. 39, pp. 21884-21889, 2015/10/07 2015, doi: 10.1021/acsami.5b06329.

[39]　J. W. Choi *et al.*, "High spectro-temporal compression on a nonlinear CMOS-chip," *Light: Science & Applications,* vol. 10, no. 1, p. 130, 2021/06/18 2021, doi: 10.1038/s41377-021-00572-z.

[40]　Y. Cao *et al.*, "Thermo-optically tunable spectral broadening in a nonlinear ultra-silicon-rich nitride Bragg grating," *Photon. Res.,* vol. 9, no. 4, pp. 596-604, 2021/04/01 2021, doi: 10.1364/PRJ.411073.

[41]　C.-L. Wu *et al.*, "Enhancing Optical Nonlinearity in a Nonstoichiometric SiN Waveguide for Cross-Wavelength All-Optical Data Processing," *ACS Photonics,* vol. 2, no. 8, pp. 1141-1154, 2015/08/19 2015, doi: 10.1021/acsphotonics.5b00192.

[42]　C, Lacava *et. al*, "Si-rich Silicon Nitride for Nonlinear Signal Processing Applications," *Sci Rep*, vol. 7, no. 22, 2017, doi: 10.1038/s41598-017-00062-6.

[43]　D. T. H. Tan, K. J. A. Ooi, and D. K. T. Ng, "Nonlinear optics on silicon-rich nitride - a high nonlinear figure of merit CMOS platform [Invited]," *Photon. Res.,* vol. 6, no. 5, pp. B50-B66, 2018/05/01 2018, doi: 10.1364/PRJ.6.000B50.

[44]　T. Hiraki, T. Aihara, H. Nishi, and T. Tsuchizawa, "Deuterated SiN/SiON Waveguides on Si Platform and Their Application to C-Band WDM Filters," *IEEE Photonics Journal,* vol. 9, no. 5, pp. 1-7, 2017, doi: 10.1109/JPHOT.2017.2731996.